# Unambiguous determination of the THz conductivity in driven $YBa_2Cu_3O_{6+x}$

K. Kaj[1], K.H. Martens[1], D. Choi[1], M. Buzzi[1], D. Nicoletti[1], Y. Liu[2], B. Keimer[2], A. Cavalleri[1,3]

[1]*Max Planck Institute for the Structure and Dynamics of Matter, Hamburg, Germany*

[2]*Max Planck Institute for Solid State Research, Stuttgart, Germany*

[3]*Clarendon Laboratory, Department of Physics, Oxford University, Oxford, UK*

**Abstract**

Mid-infrared pump, THz probe experiments in $YBa_2Cu_3O_{6+x}$ yield evidence of a high-temperature transient state of the material with the optical properties of a superconductor. However, there are ambiguities in the extracted transient *c*-axis THz optical properties, primarily resulting from the sublinear growth of the THz response with increasing mid-infrared pump fluence. It was proposed that this effect may be related to saturation of the material's response, which would then perturb its spatial profile and lead one to overestimate the induced change in the transient optical properties. Yet, sublinear growth of the signal could also arise from other effects unrelated to saturation, for example from a change in reflectivity or deposition depth for the pump pulses. Here, we remove this ambiguity by experimentally measuring the fluence-dependent complex optical constants at the 15-20 THz frequencies of the pump. We show that the observed sublinear growth is to be attributed in large part to the enhanced reflection and reduced penetration depth of the pump field at high fluences. When taking into account these findings, we confirm the original interpretation of a light-induced superconducting state both qualitatively and quantitatively.

**Main Text**

Excitation with coherent optical radiation has emerged as a powerful platform for controlling material properties with light, particularly through strong resonant driving of collective degrees of freedom at mid-infrared (mid-IR) frequencies [1–4]. For many materials, such as superconductors, the resulting non-equilibrium phases are ideally probed through their low-frequency electrodynamics using THz time-domain spectroscopy (THz-TDS) [5-14]. For example, driving the apical oxygen phonons in the cuprate $YBa_2Cu_3O_{6.48}$ (YBCO) yields superconducting-like features, including a low-frequency $1/\omega$ divergence in the imaginary part of the optical conductivity [5-9]. Figure 1(a) shows a schematic of a typical experiment, where an intense 15-20 THz pump pulse excites YBCO lattice vibrations, followed by a broadband, low-frequency THz probe. A representative reflected probe signal is shown in Fig. 1b, where the measured magnitude and phase of the field allow for extraction of the full complex optical properties of the photoinduced state.

However, because the THz probe has a longer wavelength than the pump, it typically penetrates much deeper into the material (about 20 times deeper along the c-axis of YBCO, see Fig. 1(c)), and therefore samples an inhomogeneously excited medium [5–14]. The measured response is reconstructed by solving the Fresnel equations for a multilayer medium, which requires an assumption about how the photoinduced THz optical properties vary with depth. Initially, the spatial profile of the light-induced optical changes was assumed to follow the exponentially decaying pump intensity, with a penetration depth independent of excitation fluence. This corresponds to a THz response that grows linearly with pump fluence. However, a different scaling, such as linear in the pump electric field (i.e., proportional to the square root of fluence), would lead to a different depth dependence [7–10]. Figure 1(c) compares the two profiles, with blue indicating linear scaling with fluence and red scaling with its square root.

The choice of spatial profile directly affects the extracted optical properties. Figures 1(d) and 1(e) show reconstructions using the linear-in-fluence (blue) and square-root (red) models, which yield similar spectral shapes but different magnitudes of the photoinduced conductivity. Other assumptions for the spatial profiles have also been proposed, including one based on a saturating fluence dependence that accounts for saturable absorption as the pump propagates through the sample [15]. The depth dependence of the photoinduced response is thus an intrinsic source of ambiguity in its quantitative extraction. To resolve this ambiguity, one has to determine the fluence dependence and hence the spatial profile self-consistently from the data. First the data are extracted assuming a linear pump fluence dependence of the measured induced optical change. The latter is then fitted with a power-law model leading to an updated the spatial profile and the data is extracted based on this new assumption. The process is iterated until the last assumed fluence dependence and the resulting one are self-consistent.

A second ambiguity concerns the optical properties at the pump frequency itself. The reconstruction requires one to know the optical constants at the pump frequency at each excitation fluence. These may themselves be modified by strong photoexcitation, as expected in systems with strongly coupled degrees of freedom, where nonlinear interactions such as nonlinear phononics have been observed [5–14,16–21]. A quantitative reconstruction therefore requires measuring the pump frequency complex optical response at the high fluences used in the experiments. At 15-20 THz this is challenging because resolving both the amplitude and phase requires carrier-envelope-phase (CEP) stable mid-IR pulses and electro-optic sampling with pulses of ca. 10 fs duration.

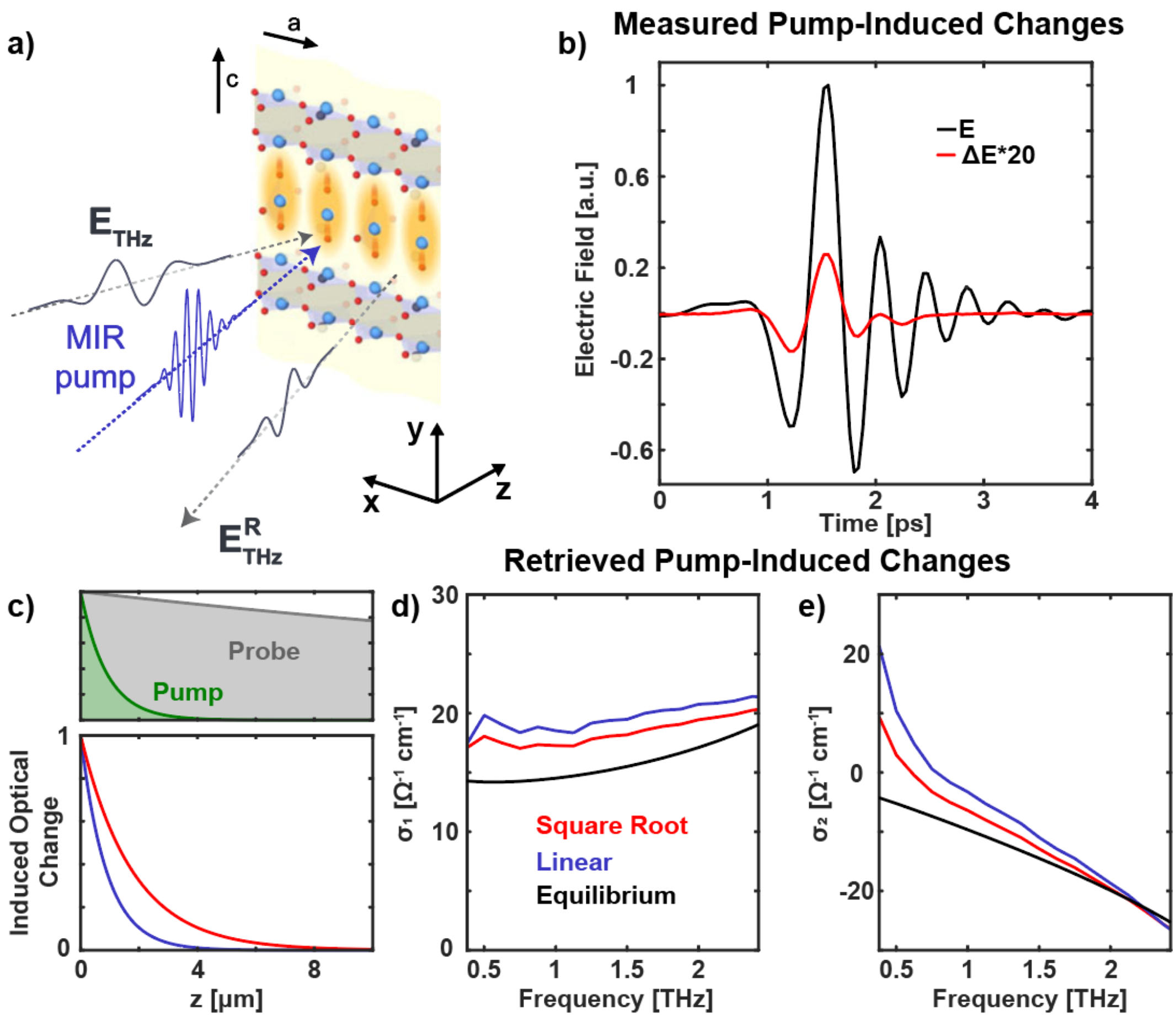

**Figure 1.** (a) Schematic of the mid-IR pump, THz probe experiment on YBCO. (b) Representative reflected probe field, showing the equilibrium field $E$ (black) and photoinduced change $\Delta E$ (red). (c) Upper: Spatial profiles for *c*-axis polarized light, the THz probe field at 1 THz (black) penetrates substantially deeper than the mid-IR pump intensity (green). Lower: Corresponding profiles of the photoinduced THz optical change for the linear (blue) and square root (red) models. (d,e) Real and imaginary parts of the equilibrium (black) and photoinduced optical conductivity reconstructed with the linear (blue) and square-root (red) models.

Here, we directly measured the pump-frequency optical response of $YBa_2Cu_3O_{6.48}$ in the same conditions of the THz experiments to investigate the origin of the sublinear growth of the near 1 THz response. Using mid-IR pump–probe time-domain spectroscopy at 15–20 THz we resolved the photoinduced changes in amplitude and phase of the reflected field, and hence the complex, fluence-dependent response of the resonantly driven apical oxygen phonons. These are the same optical parameters required to reconstruct the THz response in previous studies of light-induced superconductivity in YBCO [5–8]. We identify two effects that are important for this reconstruction.

First, the pump penetration depth decreases with increasing fluence, confining the photoinduced change to a progressively thinner layer. At 1 THz, this reduces the measured signal because the probe samples a smaller photoexcited volume. Neglecting this effect increasingly underestimates the local change required to reproduce the measured response, altering both the magnitude and fluence dependence of the reconstructed optical properties and producing "spurious evidence" for saturation.

Second, the pump reflectivity increases across the driven phonon range, so the absorbed pump fluence grows more slowly than the incident fluence. Neglecting this effect overestimates the absorbed fluence by a factor $(1 - R_0)/(1 - R(F))$, stretching the fluence ($F$) axis and potentially producing an apparent saturation. As shown below, incorporating both effects reveals a fluence dependence in the photoinduced response closer to linear.

The mid-IR pump, THz probe data analyzed here were reported in Ref. [8], where experimental details are given. The mid-IR pump–probe setup is shown in Fig. 2(a). Part of the output of a Ti:sapphire amplifier pumped two optical parametric amplifiers (OPAs), seeded by the same white-light continuum. The signal beam from each OPA was combined in GaSe to generate the mid-IR pump, while the idlers were combined in a separate GaSe to generate the probe. Common seeding renders both pulses CEP stable [22]. The probe field reflected off the YBCO sample was measured by electro-optic sampling in a third GaSe crystal using a gate pulse from a separate noncollinear OPA (NOPA), compressed with a chirped-mirror pair. This yielded the full complex photoexcited reflection coefficient.

Figures 2(b-e) summarizes the mid-IR pump–probe measurements at 100 K for a pump fluence of 7.8 $mJ/cm^2$, with pump and probe polarized along the *c*-axis, taken at the peak of the pump-probe response. These conditions match those of Ref. [8], so the extracted optical properties apply directly to the THz experiments. The dielectric function, $\varepsilon_1 + i\varepsilon_2$ (Figs 2(b) and 2(c)), is dominated by the two apical oxygen phonons, which produce peaks in $\varepsilon_2$ at their transverse-optical (TO) frequencies and the peak–dip structures in $\varepsilon_1$ [23,24]. Their strongly asymmetric line shapes in equilibrium have been described with a Fano model [23]. Photoexcitation strengthens the phonon response (red), enhancing the peaks in $\varepsilon_2$ and the peak–dip features in $\varepsilon_1$, most clearly between 18 and 19.3 THz. These are the only phonons known to induce superconducting-like THz properties when driven in YBCO, and their excitation has also been linked to coherent Josephson-plasmon oscillations observed by time-resolved second-harmonic generation [5-9,18,21]. Their nonlinear response at high fluence may therefore be directly relevant to the emergence of the photoinduced state. A detailed theoretical treatment is left for future work.

Figure 2(d,e) shows the pump-frequency optical properties that enter the reconstruction. The equilibrium and photoexcited penetration depths, shown in Fig. 2(d), both exhibit minima near the TO frequencies of the apical oxygen phonons. Photoexcitation reduces the penetration depth across 15.8-19.3 THz, most strongly near the phonon frequencies, so at high fluence the pump decays faster than predicted from the equilibrium properties. At the same time, the reflectivity increases throughout the phonon range, shown in Fig. 2(e), so a smaller fraction of the incident pump is absorbed than the equilibrium reflectivity implies. The pump fluence profile must be determined using the absorbed fluence, which is affected by the changes to the pump reflectivity.

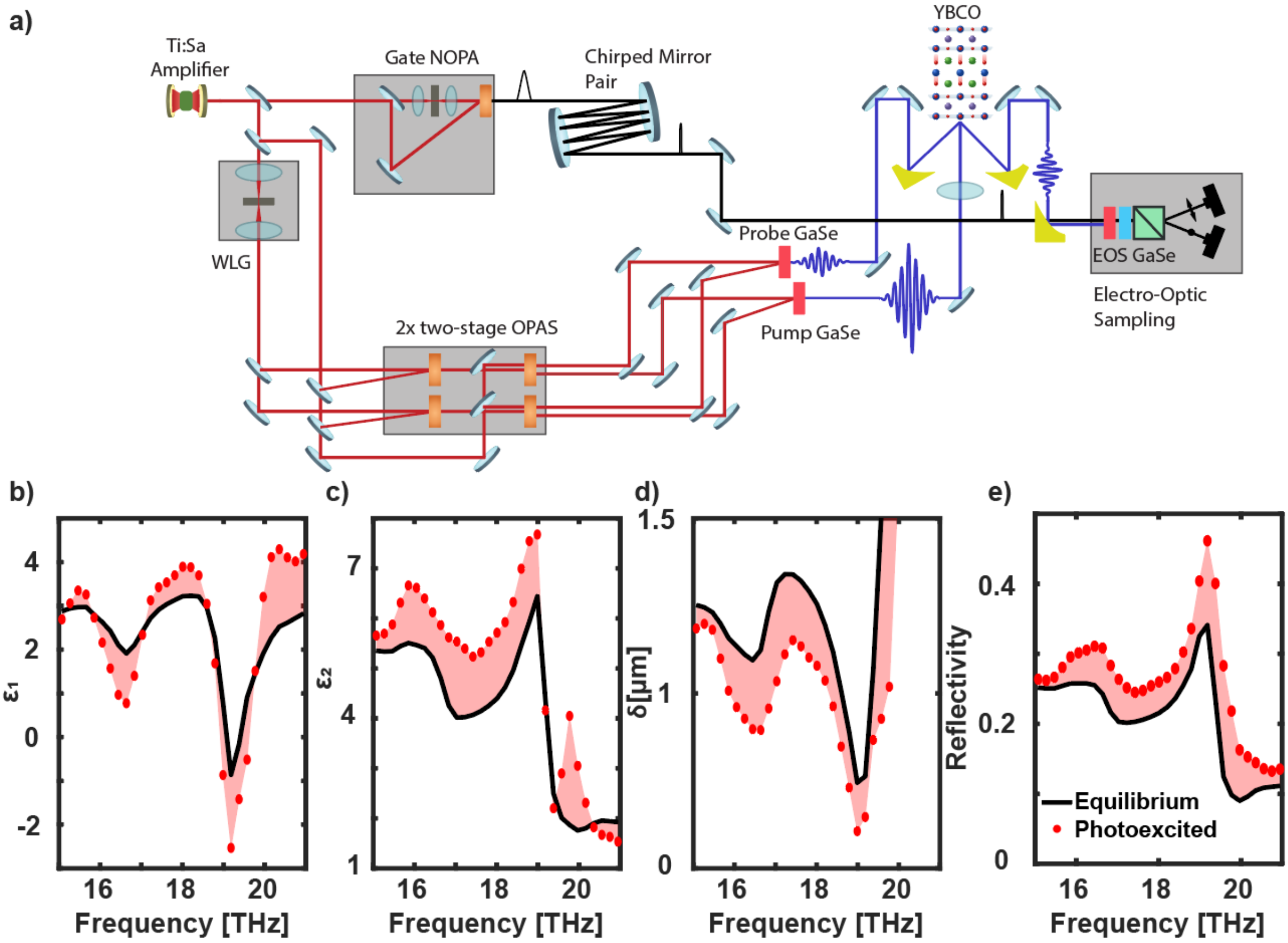

**Figure 2.** (a) Schematic of mid-IR pump–probe setup. CEP-stable pump and probe pulses are generated by difference-frequency mixing of the signal and idler outputs of two OPAs in separate GaSe crystals. The reflected probe is detected by electro-optic sampling (EOS) in a third GaSe crystal using a short gate pulse generated by a separate noncollinear OPA and compressed with a chirped-mirror pair. (b-e) Mid-IR optical properties of YBCO in equilibrium (black) and after photoexcitation (red) at 100 K with a pump fluence of 7.8 mJ/cm$^2$. (b,c) Real and imaginary part of the dielectric function, (d) penetration depth, (e) reflectivity.

Figure 3 shows the fluence dependence of the change in the penetration depth and reflectivity, averaged over each phonon resonance. The penetration depth decreases and the reflectivity increases monotonically with fluence. Both effects therefore grow in importance at high fluence and must be included in the reconstruction.

We now incorporate the measured pump-frequency optical properties into the reconstruction of the photoinduced response, removing the ambiguity in the pump parameters introduced by their fluence dependence. We apply the self-consistent iterative procedure described above, starting from a linear fluence dependence and progressively updating the spatial profile based on a power-law fit to the reconstructed response, until convergence. Since the multilayer model is formulated in terms of a spatially dependent conductivity change [25], we fit the total conductivity change, $\sum_{\omega}[|\Delta\sigma_1(\omega)| + |\Delta\sigma_2(\omega)|]$, over the 0.5–2.5 THz bandwidth. This choice is justified by the similar fluence dependences of $\Delta\sigma_1$ and $\Delta\sigma_2$ in YBCO [25].

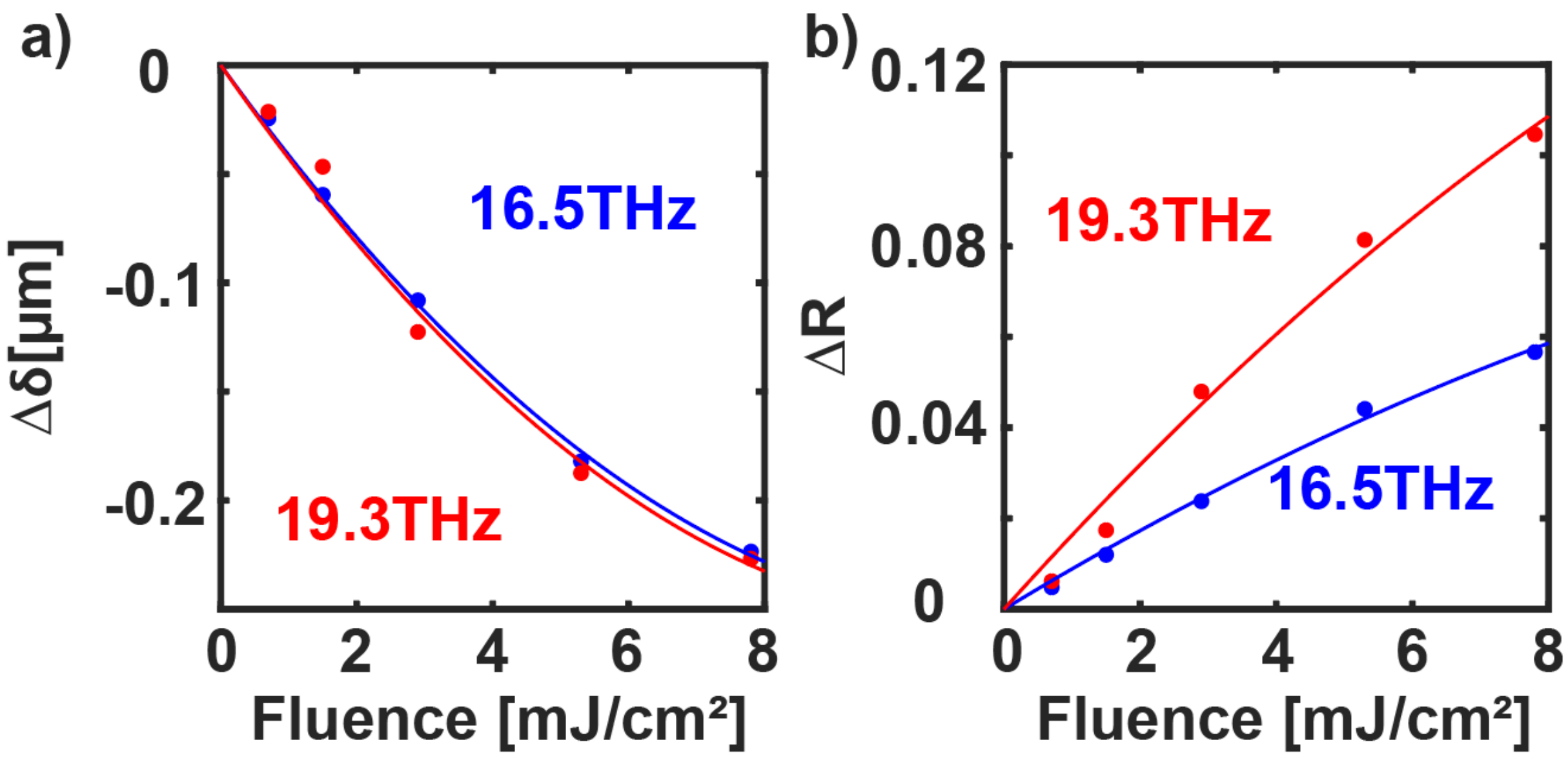

**Figure 3** Fluence dependence of the penetration depth and reflectivity changes, averaged over the 16.5 THz (blue) and 19.3 THz (red) phonon resonances.

Figure 4(a) shows the total photoinduced conductivity change for the 19.3 THz pumping as a function of absorbed fluence, reconstructed with the equilibrium (blue) and photoexcited (red) pump parameters, with solid curves showing the power-law fits. Accounting for the photoinduced changes in the optical properties at the pump frequency, the reduced penetration depth leads to a larger reconstructed conductivity change at each incident fluence, while the increased reflectivity reduces the absorbed fluence. The dependence on absorbed fluence is therefore steeper and closer to linear. The same trend is observed for 16.5 THz pump, shown in Fig. 4(d). We also tested the saturation model proposed in Ref. [15], which describes the measured fluence dependence less accurately than a power law and yields a poorer fit upon iterative reconstruction [25]. We therefore do not pursue this model further.

Figures 4(b,c) and 4(e,f) show the reconstructed conductivities for 19.3 and 16.5 THz pumping, respectively. Blue data use the conventional analysis (linear-in-fluence profile, equilibrium pump parameters), while red data include the self-consistent fluence dependence, the absorbed fluence from the measured reflectivity, and the fluence-dependent pump penetration depth. For both pump frequencies the qualitative features are robust, $\sigma_1(\omega)$ shows a nearly frequency independent increase, and $\sigma_2(\omega)$ exhibits a low-frequency $1/\omega$ divergence, consistent with the response of a dissipation-less superfluid below $T_C$ [5-8].

The self-consistent reconstruction yields a slightly smaller $1/\omega$ divergence in $\sigma_2(\omega)$ at both pump frequencies. The difference is small because the two corrections partially compensate: the sublinear fluence dependence increases the effective photoexcited volume, while the reduced pump penetration depth at high fluence counteracts this effect. Two-fluid fits give superfluid densities, $(\omega\sigma_2)_{\omega\to 0}$, that are 81% (19.3 THz) and 92% (16.5 THz) of those from the linear reconstruction. For 16.5 THz pump, the two conductivities are barely distinguishable. Unlike previous analyses, our approach determines the depth profile of the photoinduced response self-consistently and uses pump optical constants measured at the excitation fluences relevant to pump–probe experiments, rather than their equilibrium values. All input parameters are therefore either measured directly or extracted self-consistently from the data, allowing a reliable quantitative determination of the photoinduced THz conductivity, as shown in Fig. 4. This determination retains both the qualitative features and the overall magnitude of the superconducting-like response obtained from the linear analysis reported previously.

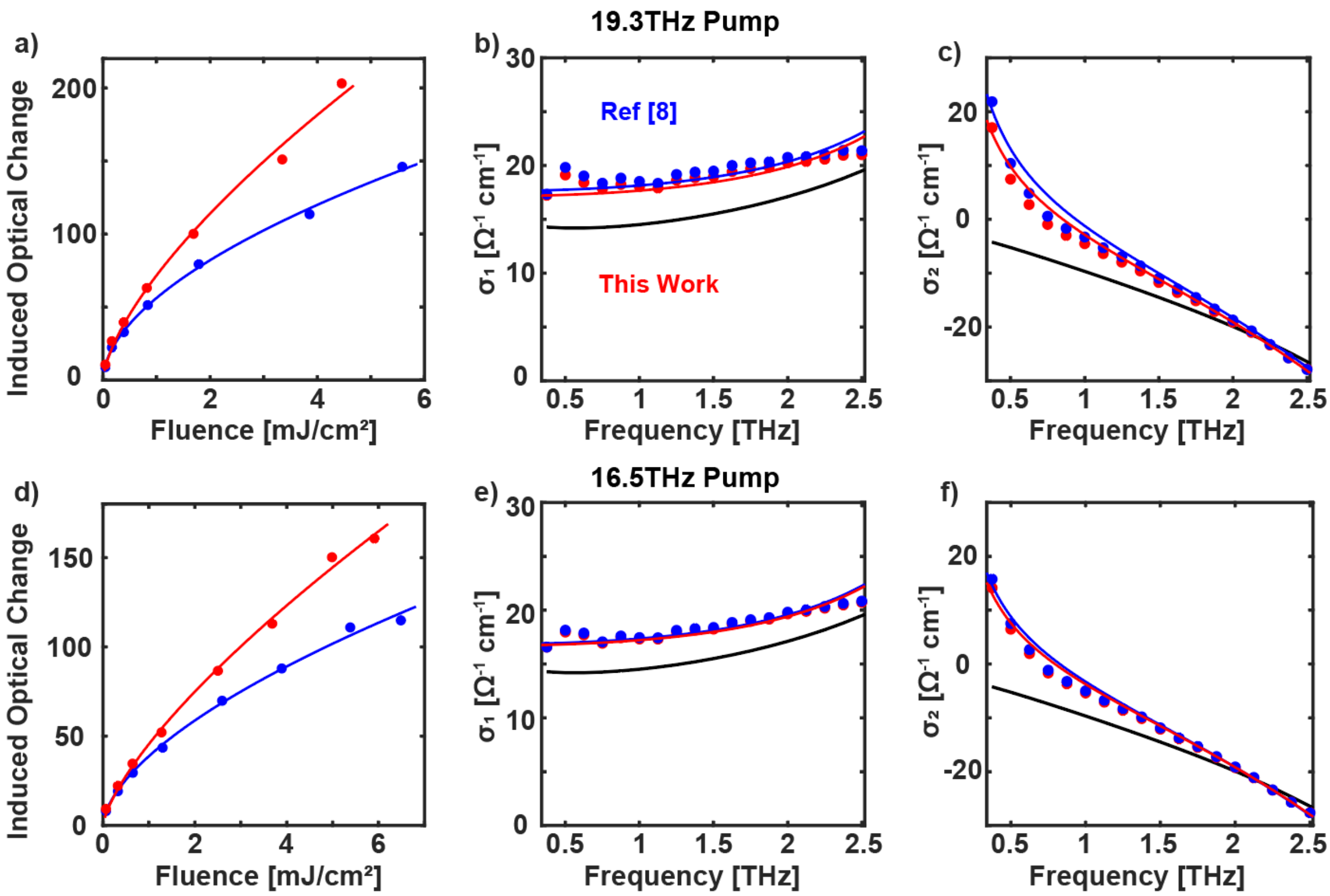

**Figure 4** Reconstruction of the photoinduced THz response for 19.3 THz (a–c) and 16.5 THz (d–f) pumping. (a) Total photoinduced conductivity change, $\sum_{\omega}[|\Delta\sigma_1(\omega)| + |\Delta\sigma_2(\omega)|]$, summed over 0.5-2.5 THz, as a function of absorbed fluence, using equilibrium (blue) and photoexcited (red) pump parameters. Solid curves are the power-law fits used in the self-consistent reconstruction. (b,c) Real and imaginary parts of the reconstructed conductivity at 100 K: conventional linear-in-fluence reconstruction with equilibrium pump parameters (blue) and self-consistent power-law reconstruction with the measured fluence-dependent pump parameters (red). Solid curves are two-fluid-model fits to the data. (d–f) Same quantities as (a–c) for 16.5 THz pumping.

We have measured the fluence-dependent mid-IR optical properties of YBCO and shown that strong photoexcitation reduces the pump penetration depth and increases its reflectivity. Combining these measurements with a self-consistent reconstruction of the photoinduced THz optical properties, we find that the characteristic $1/\omega$ divergence of the imaginary conductivity, indicative of a superfluid response below $T_C$, is preserved.

These corrections apply to any pump-probe experiment in which the probe penetrates deeper than the pump [5–8,10-13]. Fluence dependent changes in the pump penetration depth modify the photoexcited volume, and changes in pump reflectivity modify the absorbed fluence and hence the inferred fluence scaling. The latter applies even when pump and probe penetration depths are comparable. In YBCO, these effects bring the reconstructed response closer to the linear-in-fluence result, but in other materials they may substantially increase or decrease the inferred photoinduced changes. A quantitative analysis of high-fluence pump-probe experiments therefore requires knowledge of the optical response at the pump frequency.

## Supplemental Material

### S1. Experimental Methods

To determine the photoinduced changes in the mid-infrared (mid-IR) optical constants of $YBa_2Cu_3O_{6.48}$ (YBCO), we performed mid-IR pump-probe measurements. The YBCO crystal was manually polished to expose an *ac* surface so that both pump and probe pulses could be polarized along the *c*-axis. As the sample size along the *c*-axis is comparable to the pump and probe spot sizes it was glued onto a holder with a knife-edge shape. In this way, possible reflections from the parts of the pump and probe beam hitting sample holder are deflected away from the collection optics and do not interfere with the measurement. The holder was then mounted onto the cold finger of a He-cooled cryostat to control the sample temperature.

The mid-IR pump-probe setup was driven by a 1-kHz repetition rate Ti:sapphire regenerative amplifier delivering 4.5-mJ, 100-fs pulses at 800 nm. Most of the pulse energy was used to pump two 2-stage optical parametric amplifiers (OPAs) seeded by the same white-light continuum. The two signal beams were mixed in a 600-µm thick GaSe crystal to generate the pump pulses, while the idler beams, separated from the signals with dichroic mirrors, were mixed in a second 600-µm GaSe crystal to generate the probe pulses. Because both OPAs share a common seed, the pump and probe pulses are carrier-envelop-phase (CEP) stable. The pump fluence was adjusted with a pair of wire-grid polarizers, and the probe was attenuated with a second polarizer pair to keep its fluence low and ensure linear response. The pump was focused onto the sample at normal incidence to a spot size of 180 µm FWHM, and the probe at $30^{o}$ incidence to a spot size with 140 µm FWHM major axis. Both beams were polarized along the *c*-axis, with the probe *s*-polarized.

The photoinduced changes in the reflected probe field were measured by mid-IR time-domain spectroscopy. The reflected probe was electro-optically sampled in a 100-µm thick GaSe crystal using a broadband gate pulse that was generated in a separate noncollinear OPA (NOPA) and compressed with a chirped mirror pair. Pump-probe traces were acquired following the scheme of Ref. [1], keeping fixed the delay $\tau$ between the gate and pump pulses while varying the delay $t$ of the probe pulse, so that each point of the reflected field waveform was sampled at the same delay after photoexcitation. The equilibrium reflected field, $E_R(t)$, and its photoinduced change, $\Delta E_R(t,\tau)$, were recorded simultaneously by chopping the pump and probe beams at different frequencies and acquiring the electro-optic sampling signal shot by shot with a data-acquisition card, following a similar scheme as what reported in Refs. [2,3].

The transient THz optical properties were instead reconstructed starting from the raw data used for the study reported in Ref. [4]. The photo-excitation conditions used for the mid-IR pump, mid-IR probe experiment presented here are similar to those in Ref. [4].

### S2. Reconstruction of the Mid-Infrared Transient Optical Properties

The equilibrium reflected field $E_R(t)$ and its photoinduced change $\Delta E_R(t,\tau)$ were then processed to obtain the photo-excited complex reflection coefficient of the sample $\tilde{r}(\omega,\tau)$ by taking their Fourier transform along $t$ and using the following equation:

$$\frac{\Delta\tilde{E}_R(\omega,\tau)}{\tilde{E}_R(\omega)} = \frac{\tilde{r}(\omega,\tau)-\tilde{r}_0(\omega)}{\tilde{r}_0(\omega)}$$

where $\tilde{r}_0(\omega)$ is the equilibrium complex reflection coefficient, obtained from equilibrium Fourier Transform Infrared Spectroscopy measurements [5,6]. As discussed in the main text, in the THz experiments described here the pump penetrates less than the probe, which interrogates an inhomogeneously excited volume (Fig. 1(b), main text). This is taken into account by "slicing" the probed thickness along the depth coordinate $z$ into thin layers where we assume that the pump-induced changes in the optical conductivity scale according to the pump intensity in the layer, *i.e.* $\tilde{\sigma}(\omega, z, \tau) = \tilde{\sigma}_0(\omega) + \Delta\tilde{\sigma}(\omega, \tau, I(z))$. The pump intensity is assumed to follow the dependence $I(z) = I_0 e^{-z/\delta_{pump}}$, where $\delta_{pump} = \lambda_{pump}/4\pi\, \mathrm{Im}(n(\omega_{pump}))$. To avoid making assumption on how $\Delta\tilde{\sigma}(\omega, \tau)$ scales with the local pump fluence we implement a self-consistent procedure. First the data is reconstructed assuming that $\Delta\tilde{\sigma} \propto I(z)$. This yields fluence-dependent conductivity data from which the total frequency-integrated photoinduced conductivity change, $\Delta\sigma = \sum_\omega[|\Delta\sigma_1(\omega)| + |\Delta\sigma_2(\omega)|]$ can be obtained. These data are then fitted with a power law function (*i.e.*, $\Delta\sigma \propto I(z)^n$) and the reconstruction is run again with a new guess for the fluence dependence (*i.e.*, $\Delta\sigma \propto I(z)^n$) obtained from the previous reconstruction. The process is then iterated until the value of the exponent $n$ does not change between two consecutive steps. The final results are shown in Fig. 2 and Fig. 3 of the main text.

**S3. Reconstruction of the THz transient Optical Properties**

A procedure similar to the one described for the extraction of the transient mid-IR optical properties (see Supplemental Section S2) was used to extract the THz range. Here, the values of the fluence dependent pump penetration depth $\delta_{pump}$ were used to determine the pump intensity as a function of depth according to $I(z) = I_0 e^{-z/\delta_{pump}}$. The pump induced changes were then assumed to scale following a power law $\Delta\tilde{\sigma} \propto I(z)^n$ and the exponent was determined self-consistently as described in Supplemental Section S2. The final results are shown in Fig. 4 of the main text.

Figure S2 shows the coefficient of determination, $R^2$, of the fluence dependence fits at each iteration step for the power law model and an additional model that follows a saturation behavior discussed in Ref [7]. With the power law, the procedure converges within one or two iterations for both pump frequencies, and the fit quality is unchanged. The saturation model also converges to a self-consistent set of parameters, meaning the parameters obtained from fitting the fluence dependence match those used to construct the spatial profile. However, its fit quality decreases with each iteration, so the converged solution describes the data less well than the initial fit. This provides further evidence that the saturation model does not describe the photoinduced response of YBCO.

To gauge the effect that the pump-induced changes in the mid-IR optical properties carry on the fluence dependence without accounting for the self-consistent process, Figure S2 shows the fluence dependence obtained as initial step of the reconstruction. Here, the depth profile of the changes in the optical constants is modeled as an exponential decay whose characteristic length is the pump penetration depth, which in equilibrium is 1.1 µm at 16.5 THz and 0.89 µm at 19.3 THz. Figures S1(a) and S1(c) show the results obtained with these equilibrium parameters, fitted with either a power law or a saturation model [7]. For both pump frequencies, the power law describes the data better than the saturation model.

Including the changes of the optical response functions at the pump frequency modifies this fluence dependence in two ways. First, the pump penetration depth decreases with increasing fluence, so the photoinduced change reconstructed at high fluence is larger than that obtained with the equilibrium penetration depth. Second, the increase in pump reflectivity reduces the absorbed fluence assigned to each measurement, with the effect becoming more pronounced at higher fluences. Both effects steepen the fluence dependence of the total photoinduced change, shown in Figs. S1(b) and S1(d), raising the power law exponent towards unity and increasing the saturation fluence, $F_{sat}$, obtained from the saturation model fit. All values of $F_{sat}$ are given in terms of absorbed fluence.

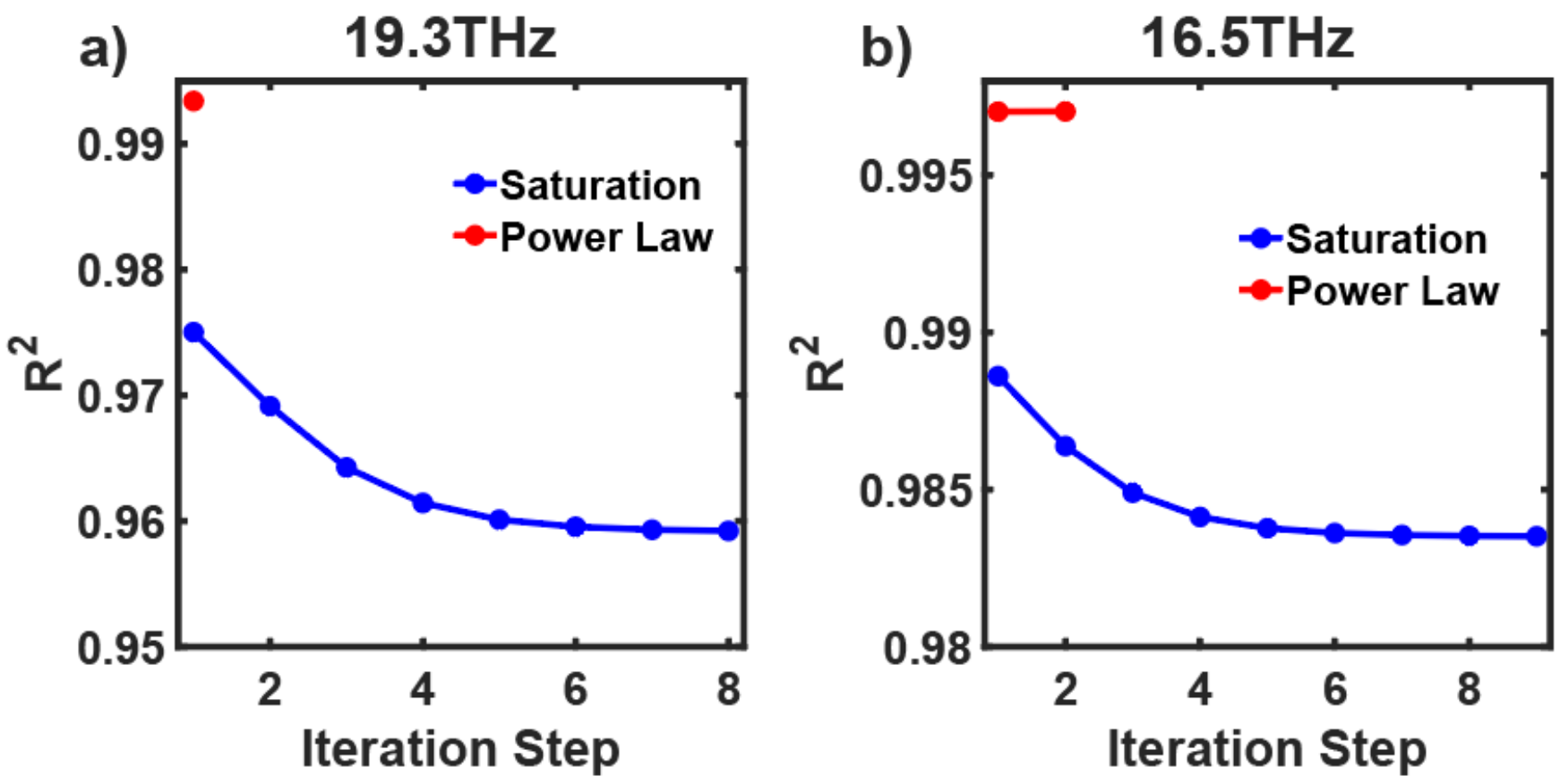

**Figure S1.** Coefficient of determination, $R^2$, of the fits to the fluence dependence of the total photoinduced conductivity change at each iteration step of the self-consistent reconstruction, for (a) 19.3 THz and (b) 16.5 THz pumping, using the power-law and saturation models. The fit quality of the saturation model decreases with iteration.

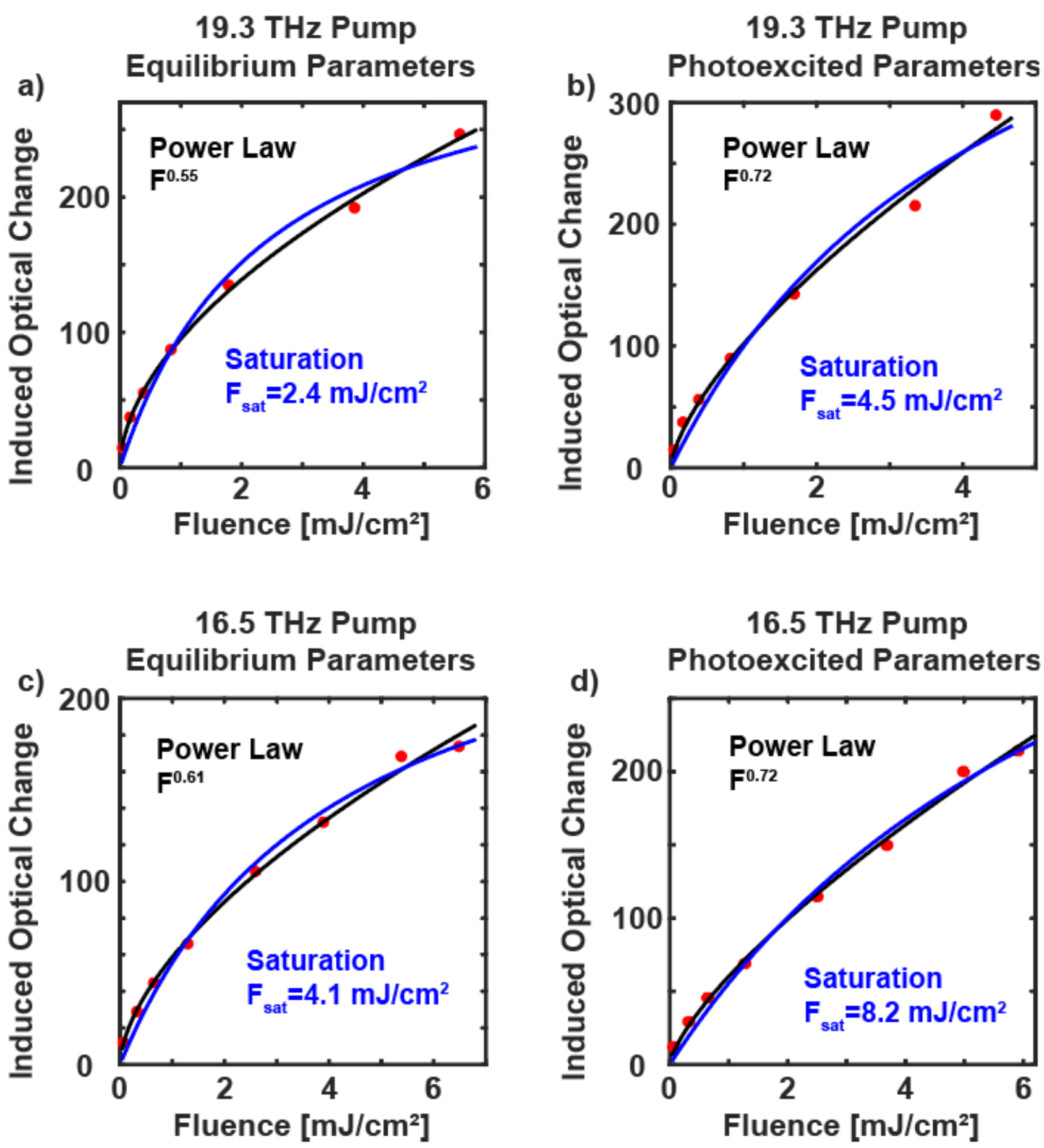

**Figure S2.** Total frequency-integrated photoinduced conductivity change, $\sum_{\omega}[|\Delta\sigma_1(\omega)| + |\Delta\sigma_2(\omega)|]$, from the initial linear-in-fluence reconstruction as a function of absorbed fluence. (a,c) Reconstruction using the equilibrium pump optical constants for 19.3 THz and 16.5 THz pump frequency respectively. (b,d) Reconstruction using the photoexcited pump optical constants measured in this work. Black curves are power law fits and blue curves are saturation model fits.

## S4. Choice of optical quantity for the reconstruction

In principle, any of the equivalent optical response functions (complex optical conductivity, refractive index, or dielectric function) can be used to reconstruct the photoinduced THz properties. Here, all reconstructions are performed directly in terms of the complex optical conductivity, $\tilde{\sigma}(\omega) = \sigma_1(\omega) + i\sigma_2(\omega)$. Because a single depth profile is applied to the complex photoinduced change, an appropriate choice requires that the real and imaginary parts exhibit similar fluence dependence. Figure S3 shows the fluence dependence of $\sum_\omega |\Delta\sigma_1(\omega)|$ and $\sum_\omega |\Delta\sigma_2(\omega)|$, the photoinduced changes in the real and imaginary parts of the optical conductivity, integrated over the 0.5-2.5 THz bandwidth. Both follow similar power-law scaling, so the optical conductivity is a suitable quantity for the reconstruction.

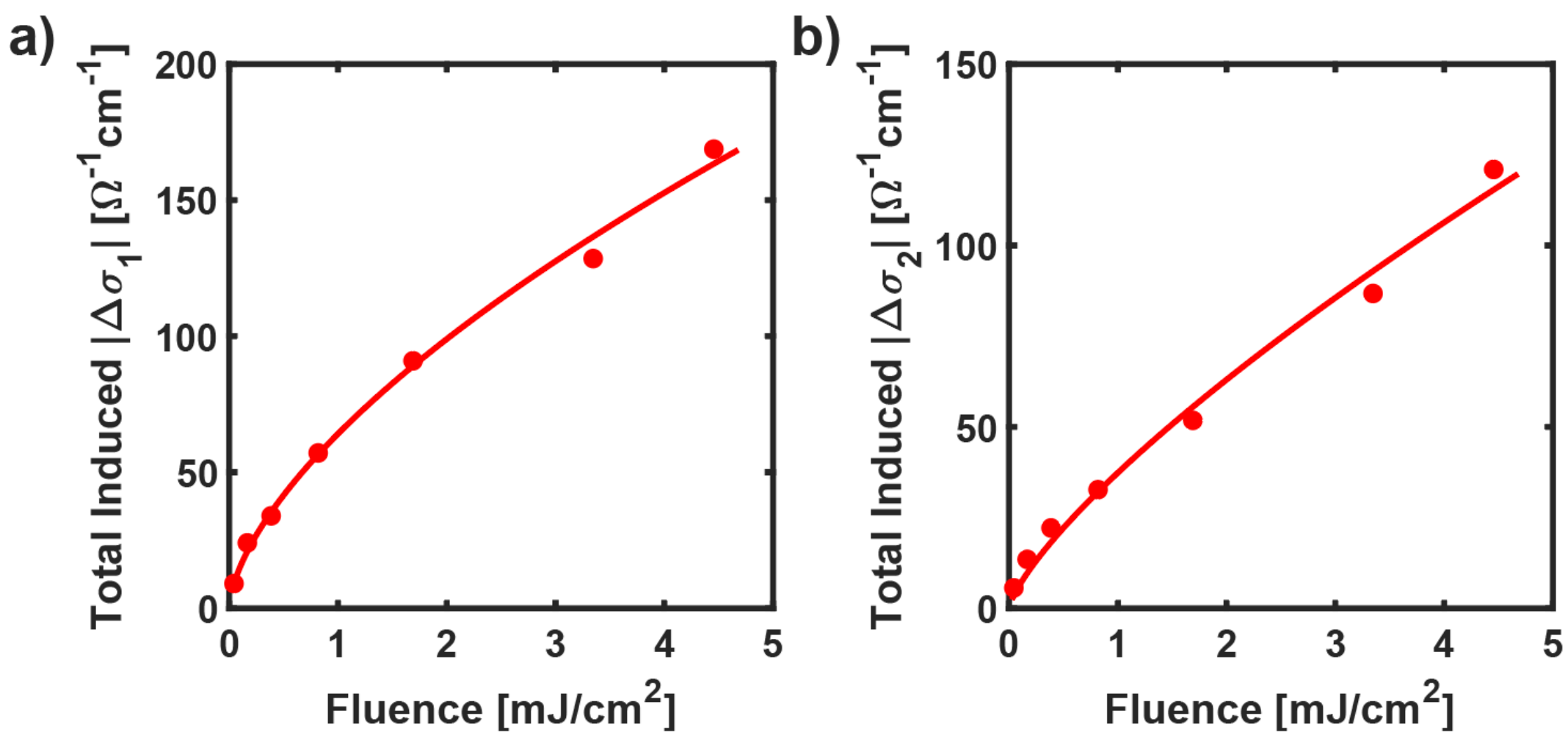

**Figure S3.** Fluence dependence of the photoinduced change in the (a) real and (b) imaginary parts of the optical conductivity, summed over 0.5-2.5 THz. Both exhibit similar power-law scaling.